\documentstyle[prl,twocolumn,epsf,floats,aps]{revtex}

\begin{document}
\draft

\twocolumn[\hsize\textwidth\columnwidth\hsize\csname
@twocolumnfalse\endcsname

\title{Nonabelian braid statistics versus projective permutation
statistics}
\author{N. Read}
\address{Department of Physics, Yale University, P.O. Box 208120,
New Haven, CT 06520-8120}
\date{\today}
\maketitle

\begin{abstract}
Recent papers by Finkelstein, Galiautdinov, and coworkers {[J.
Math. Phys. {\bf 42}, 1489, 3299 (2001)]} discuss a suggestion by
Wilczek that nonabelian projective representations of the
permutation group can be used as a new type of particle
statistics, valid in any dimension. Wilczek's suggestion was based
in part on an analysis by Nayak and Wilczek (NW) of the nonabelian
representation of the braid group in a quantum Hall system. We
point out that projective permutation statistics is not possible
in a local quantum field theory as it violates locality, and show
that the NW braid group representation is not equivalent to a
projective representation of the permutation group. The structure
of the finite image of the braid group in a
$2^{n/2-1}$-dimensional representation is obtained.
\end{abstract}

%\pacs{PACS numbers:}
\vspace{0.1in} ]

%\newpage

Moore and Read \cite{mr} showed that a physical realization of
nonabelian statistics (as a nonabelian representation of the braid
group) was a possibility in a quantum Hall effect system. The
Moore-Read state now seems likely to be the ground state in the
$\nu=5/2$ quantum Hall effect (for a review, see \cite{read00}).
The nonabelian statistics was analyzed further \cite{nayak,rr,rg},
and in particular Nayak and Wilczek (NW) \cite{nayak} showed that
exchange of the quasiparticles by braiding can be represented
using a subgroup of the rotation group SO($n$), acting in the
spinor (projective) representation, using Clifford algebra
methods. Wilczek \cite{wilczek} then proposed a connection with
the projective representations of the permutation group, and
suggested that such ``projective permutation statistics'' are a
possibility in any space dimension. This was explored extensively
in Refs.\ \cite{bfgs,fg,g}, where it was termed ``Clifford
statistics''. In view of the interest in nonabelian statistics
also in connection with quantum computation \cite{freed}, it seems
worthwhile to correct the confusion that has arisen.

To begin, consider $n$ indistinguishable point objects in a
two-dimensional plane. For generic positions, they can be
projected onto a generic line in such a way that they do not
coincide, and can then be labelled $1$, \ldots, $n$ in sequence
from left to right. The permutation (or symmetric) group acting on
the objects is generated by the set of $s_j$, $j=1$, \ldots,
$n-1$, that exchange objects $j$, $j+1$.
The generators obey relations%
\begin{eqnarray}
s_j^2&=&1,\\(s_js_k)^3&=&1\qquad(|j-k|=1),\label{symmrel2}\\
s_js_k&=&s_ks_j\qquad(|j-k|>1) \label{symmrel3}
\end{eqnarray}
(where $1$ denotes the identity element of the group), and this
set of generators and relations defines the symmetric group on $n$
objects $S_n$. It has $n!$ elements.

Similarly, the braid group $B_n$ is generated by nearest-neighbor
transpositions $t_j$, but now they do not square to the identity.
The braid group can be defined by the relations (see e.g.\
\cite{yswu})%
\begin{eqnarray}
t_jt_kt_j&=&t_kt_jt_k\qquad(|j-k|=1),\label{braidrel1}\\
t_jt_k&=&t_kt_j\qquad(|j-k|>1) \label{braidrel2}.
\end{eqnarray}
The braid group is an infinite discrete group.

The projective representations of the symmetric group $S_n$ can be
viewed as ordinary linear representations of a covering group,
that is a nontrivial central extension of $S_n$ by U($1$) (or by a
subgroup thereof). The central extensions of $S_n$ by U($1$) are
classified up to isomorphism by the cohomology group $H^2(S_n,{\rm
U}(1))$ which for $n\geq 4$ is $\cong {\bf Z}_2$
\cite{weib,dijk,hh}. Hence for $n\geq 4$ (we consider only this
range from here on) there are nontrivial extensions of $S_n$ by
${\bf Z}_2$, which have $2\cdot n!$ elements, and we denote one of
these by $\widetilde{S}_n$. $\widetilde{S}_n$ can be defined by
$n$ generators $\sigma_j$ ($j=1$, \ldots, $n-1$), $z$ and
relations \cite{hh,hamer}
\begin{eqnarray}
z^2&=&1,\\ z\sigma_j&=&\sigma_jz,\\ \sigma_j^2&=&z,\label{doubrel3}\\
\sigma_j\sigma_k\sigma_j&=&\sigma_k\sigma_j\sigma_k\qquad(|j-k|=1),
\label{doubrel4}\\
\sigma_j\sigma_k&=&z\sigma_k\sigma_j\qquad(|j-k|>1).
\label{doubrel5}
\end{eqnarray}
Thus $z$ is a central element that commutes with all elements, and
can be set to either $+1$ or $-1$ in any irreducible
representation (note that we do not distinguish between the
abstract generators $z$, $s_j$, $t_j$, $\sigma_j$, etc, and their
matrix representatives in a particular representation). The
relations are the same as for the symmetric group, modulo factors
of $z$. Representations in which $z=1$ descend to linear
representations of the quotient group,
$\widetilde{S}_n/\{1,z\}\cong S_n$, while representations in which
$z=-1$ are projective representations of $S_n$. (The only other
nontrivial double cover $\widehat{S}_n$, not isomorphic to
$\widetilde{S}_n$ except for $n=6$ \cite{hh}, is obtained by using
instead generators $\sigma_j'$ which obey similar relations but
with $1$ in place of $z$ in eq.\ (\ref{doubrel3}) \cite{hh,hamer}.
In a representation in which $z=-1$, this results from setting
$\sigma_j'=i\sigma_j$ for all $j$. These were the relations used
in Ref.\ \cite{wilczek}.)

The proposal for projective permutation statistics \cite{wilczek}
was that, as quantum mechanics welcomes the use of projective
representations of symmetries, identical particles might be
described by projective representations of the permutation group.
Since the permutations do not refer to the topology of space
(unlike the braiding operations), this proposal, if correct, could
be used in any dimension (the ordering of the particles along the
line is then arbitrary). Then the operation of exchange of nearest
neighbors would be represented by an element $T_j$ acting on
Hilbert space, and in the projective permutation statistics
proposal, each $T_j$ must be either $\sigma_j$ or $-\sigma_j$,
since these are the elements that project to transpositions $s_j$
in the quotient group $S_n$. In particular, there is a
representation of $\widetilde{S}_n$ of dimension $2^{[(n-1)/2]}$
(where $[x]$ denotes the largest integer $\leq x$). This coincides
with the dimension of the representation of the braid group
identified \cite{mr,nayak,rr} in the Moore-Read quantum Hall
state, and Wilczek \cite{wilczek} claimed that this representation
of $\widetilde{S}_n$ is equivalent to the representation of the
braid group obtained in Ref.\ \cite{nayak}, up to some phase
factors that we will discuss in a moment. Note that the complex
Clifford algebra on $m$ generators $\gamma_j$, with relations
$\gamma_j^2=1$, $\gamma_j\gamma_k=-\gamma_k\gamma_j$ ($j\neq k$),
has dimension $2^m$. For $m$ even, the Clifford algebra is
isomorphic to the algebra of matrices on a vector space of
dimension $2^{m/2}$. This applies here with $m=2[(n-1)/2]$.

The difficulty with the general proposal is that statistics of
particles in quantum field theory or many-body theory must obey
locality. That is, the underlying physics is presumed to be given
by a local Hamiltonian containing local interactions between local
fields (for example, the electrons in the quantum Hall system).
The locality assumption plays a crucial role in the general
rigorous analysis of particle statistics; see e.g.\ Refs.\
\cite{dhr,bf,frs,fgm,dr,haag}. In particular, it appears that
projective representations of the permutation group are explicitly
ruled out (see e.g.\ Thm.\ 2.2.3c in Ch.\ IV of Ref.\ \cite{haag}
for the case of relativistic theories in space dimension $\geq 3$
under some technical assumptions that are relaxed by the end of
Sec.\ IV.3.3, and Sec.\ IV.5 of Ref.\ \cite{haag} for some
discussion of space dimension $2$ where the braid group enters).
The central step of the analysis is to move particles around {\em
continuously} in spacetime, and the results depend only on the
homotopy class of the path taken in configuration space. In
particular, exchanges of disjoint well-separated pairs of
particles must commute as the two orderings of the exchanges are
homotopically equivalent, so in particular $T_jT_k=T_kT_j$ for
$|j-k|>1$, or in other words the group-theoretic commutator
$T_jT_kT_j^{-1}T_k^{-1}=1$. In the projective representations of
the symmetric group, the commutator is instead $-1$ (whatever the
choice of the lift, $T_j=\sigma_j$ or $-\sigma_j$, of each $s_j$),
and so {\em projective statistics violates locality}. On the other
hand, locality is not violated by braid statistics, where
$T_j=t_j$ in some representation of $B_n$, and it is known that
nonabelian braid statistics can be realized in a local theory in
$2+1$ dimensions \cite{fgm,frs}, for example in pure Chern-Simons
gauge theory.

Independent of the physical requirement of locality, the
difference between the commutators of generators in $B_n$
(\ref{braidrel2}) and in $\widetilde{S}_n$ (\ref{doubrel5})
implies that a projective representation of $S_n$ (in which
$z=-1$) {\em cannot} also be a representation of the braid group
$B_n$, in contradiction to Wilczek's claim \cite{wilczek}. Put
another way, the image of the braid group in U($2^{[(n-1)/2]}$)
given by the representation matrices (the existence of which will
be checked later) and that of $\widetilde{S}_n$ are not isomorphic
as groups (given the way that both project to the symmetric
group). (Later we will see that these two groups, though both
finite, are actually of different orders.)

No escape from these conclusions can be found in a remark by
Wilczek \cite{wilczek} that in the quantum Hall example, the
projective statistics is combined with anyonic phase factors,
$e^{2\pi i/8}$ in a $T_j$. If this is taken to mean that the
physical exchanges $T_j$ act in a tensor product of the
$2^{[(n-1)/2]}$ dimensional representation of $\widetilde{S}_n$ as
above, with an abelian representation of the braid group
$t_j=e^{i\theta}$ for some real $\theta$, so $T_j=\sigma_j\otimes
t_j$, then it is clear that this does not affect the
noncommutation of disjoint exchanges, $T_jT_kT_j^{-1}T_k^{-1}=-1$.
(A special case is $e^{i\theta}=i$, discussed earlier.) These
generators clearly obey the relations (reintroducing $z$ for
convenience)
\begin{eqnarray}
z^2&=&1,\\ z\tau_j&=&\tau_jz,\\
\tau_j\tau_k\tau_j&=&\tau_k\tau_j\tau_k
\qquad(|j-k|=1),\\
\tau_j\tau_k&=&z\tau_k\tau_j\qquad(|j-k|>1).
\end{eqnarray}
The existence of representations $T_j=\tau_j$ of these relations
implies their consistency, and hence the existence of a nontrivial
central extension $\widetilde{B}_n$ of the braid group, defined
abstractly by the generators $z$, $\tau_j$ and the above
relations. Any of the four groups mentioned earlier,
$\widetilde{S}_n$, $B_n$, $S_n$, or $\widehat{S}_n$ can be
obtained from $\widetilde{B}_n$ by imposing additional relations
$\tau_j^2=z$, $z=1$, both of these, or $\tau_j^2=1$, respectively.
Similarly, if $t_j^{(1)}$ and $t_j^{(2)}$, $j=1$, \ldots, $n-1$,
are two representations of the braid group $B_n$, then
$t_j=t_j^{(1)}\otimes t_j^{(2)}$ gives another one. In particular,
$t_j^{(2)}=e^{i\theta}$ (for all $j$) is a one-dimensional
representation, and so a continuum of distinct representations of
the same dimension can be found for each choice of $t_j^{(1)}$'s.
In quantum Hall effect systems, such abelian tensor factors are
common, as there is a contribution to $T_j$ from the charge
degrees of freedom, which produces a $\theta$ that depends on the
filling factor.

If one considers representations modulo phase factors, then this
distinction between the commutators (\ref{braidrel2}),
(\ref{doubrel5}) cannot be made. This is the notion of isomorphism
of groups modulo scalars, in contrast to the usual isomorphism we
have been invoking so far. Isomorphism modulo scalars amounts to
isomorphism of the images of the group(s) in the projective linear
group PGL($N$) $\cong {\rm GL}(N)/{\rm GL}(1)$, or since we are
considering unitary representations, PU($N$) $\cong$
U($N$)/U($1$). However, isomorphism modulo scalars is generally
too weak a property to use in quantum physics. That is because we
must keep track of interference between processes that correspond
to distinct group operations, and the phases involved may be
relative phases that affect such interference. That is, we are
interested in more than just the representation of a group. For
example, $S_n$ has two one-dimensional representations, one in
which $s_j=+1$, one in which $s_j=-1$, corresponding to Bose and
Fermi statistics, respectively. Modulo scalars, these are
isomorphic, but linearly (and physically) they are not.

We now examine the construction of NW \cite{nayak} to find the
structure of their braid group representation of dimension
$2^{[(n-1)/2]}=2^{n/2-1}$ (we consider only $n$ even from here on;
there are similar results for $n$ odd). Essentially the same
construction, based on the Temperley-Lieb (TL) algebra specialized
to the Ising model, was obtained much earlier by Jones
\cite{jones}. See also Ref. \cite{ivanov}. NW deduce most of its
properties from the properties of conformal blocks of spin fields
in the Ising model, as in Ref.\ \cite{mr}. The central idea is
that each object corresponds to an orthogonal direction in real
$n$-dimensional Euclidean space ${\bf R}^n$, and the elementary
transpositions $T_j$ correspond to a rotation $\upsilon_j$ by
$\pi/2$ in the plane spanned by objects $j$, $j+1$, acting in one
of the two inequivalent spinor representations of dimension
$2^{n/2-1}$ of the covering group Spin($n$) of SO($n$), up to a
$j$-independent phase factor as just discussed:
$T_j=e^{i\theta}\upsilon_j$. Clearly these operations have the
effect of permuting the $n$ axes (if we ignore the direction along
each axis), and thus do project to the action of the permutation
group as desired. Each rotation can be defined as $\upsilon_j=\exp
[i(\pi/2)e_{j,j+1}]$, where $e_{j,k}$ ($j<k$) is the element of
the Lie algebra so($n$) that generates a rotation in the $jk$
plane, acting here in the chosen spinor representation. Since the
generators $e_{j,k}$ for disjoint pairs $j_1k_1$, $j_2k_2$
commute, and this remains true in any representation including the
spinors (there are no nontrivial central extensions of any
semisimple Lie algebra!), the $\upsilon_j$'s commute,
$\upsilon_j\upsilon_k\upsilon_j^{-1}\upsilon_k^{-1}=1$ for
$|j-k|>1$. Hence there is no difficulty with locality of the
proposal of Ref.\ \cite{nayak}, and so far it is consistent with
the claim that the $\upsilon_j$'s form a linear representation of
the braid group, with $t_j=\upsilon_j$. It remains to check the
other relation (\ref{braidrel1}).

To understand the structure of the representation of the braid
group of dimension $2^{n/2-1}$ given by $t_j=\upsilon_j$, it is
useful first to consider the geometry of the group of rotations by
$\pi/2$ about the axes in ${\bf R}^n$ in more detail. This amounts
to studying the group generated by elements $u_j=\exp
[i(\pi/2)e_{j,j+1}]$, where this time $e_{j,k}$ act in the
defining $n$-dimensional representation of SO($n$). The operation
$u_1$, for example, sends the point with coordinates
$(x_1,\ldots,x_n)$ to $(-x_2,x_1,x_3,\ldots,x_n)$. The group
generated by the $u_j$'s can be seen to be the set of all
permutations of $x_1$, \ldots, $x_n$, together with sign changes,
but with the condition that an even permutation is combined with
an even number of sign changes, and an odd permutation with an odd
number of sign changes. If the latter condition is dropped, we
obtain the group of all permutations and sign changes, which is
generated by all reflections in the diagonals $x_j=x_k$ ($1\leq
j<k\leq n$) and in the coordinate planes $x_j=0$, $j=1$, \ldots
$n$. This is therefore a Coxeter group, denoted ${\cal B}_n$
\cite{hump} [it is the Weyl group of so($2n+1$) and sp($2n$)]. It
can be described by generators and relations, but we will not need
these here. There is a subgroup of index $2$, which we denote
${\cal B}_n^+$, consisting of the elements that are proper
rotations, and it is exactly the group generated by the $u_j$'s.
${\cal B}_n$ is a semidirect product of $S_n$ with the group of
sign changes $({\bf Z}_2)^n$, and has order $2^n\cdot n!$. Its
rotation subgroup ${\cal B}_n^+$ has order $2^{n-1}\cdot n!$, and
is an extension of $S_n$ by $({\bf Z}_2)^{n-1}$, but not a
semidirect product (that is, there is no $S_n$ subgroup of ${\cal
B}_n^+$ that projects onto $S_n$ under the quotient map ${\cal
B}_n^+\to{\cal B}_n^+/({\bf Z}_2)^{n-1}\cong S_n$). Finally, the
cover Spin($n$) of SO($n$), and the inclusion of ${\cal B}_n^+$ in
SO($n$), induce a double cover $\widetilde{\cal B}_n^+$ (there is
a similar double cover $\widetilde{\cal B}_n$ of ${\cal B}_n$).
$\widetilde{\cal B}_n^+$, which has order $2^n\cdot n!$, is almost
the group we need. It is generated by the lifts of the $u_j$'s,
and the irreducible representations of dimension $2^{n/2-1}$ of
Spin($n$) induce representations of the same dimension of
$\widetilde{\cal B}_n^+$, which can also be viewed as projective
representations of ${\cal B}_n^+$. To find the order of the image
of $\widetilde{\cal B}_n^+$ in the irreducible spinor
representations, we note that, for $n\geq 6$, the only normal
subgroups of Spin($n$) are contained in its center, which is ${\bf
Z}_4$ ($n/2$ odd), ${\bf Z}_2\times {\bf Z}_2$ ($n/2$ even), so
the kernel of the map $\widetilde{\cal B}_n^+ \to $ U($2^{n/2-1}$)
must also be contained in the center of Spin($n$). Hence the order
of the image of $\widetilde{\cal B}_n^+$ is the same as the order
of $\widetilde{\cal B}_n^+$, within a factor of $2$ or $4$. For
$n=4$, Spin($4$) $\cong$ SU($2$)$\times$SU($2$), and the
irreducible spinor representations do not faithfully represent the
Lie algebra so($4$), so the factor could be larger.

For ${\cal B}_n^+$, it is easy to show that setting $t_j=u_j$ does
satisfy relation (\ref{braidrel1}) defining the braid group $B_n$.
To study the other groups explicitly, we resort to Clifford
algebra methods. The reducible spinor representation of so($n$),
of dimension $2^{n/2}$, can be naturally constructed as a
representation of the even part of a complex Clifford algebra on
$n$ generators by setting $e_{j,k}=-i\gamma_j\gamma_k/2$. The
representation splits into two irreducibles of dimension
$2^{n/2-1}$ (this is also the structure of the Temperley-Lieb
algebra in the Ising model \cite{jones}, and of a full Clifford
algebra on only $n-1$ generators, which Jones constructs
\cite{jones}). Spin($n$) and its center (and hence
$\widetilde{\cal B}_n^+$, by a similar argument to that in the
previous paragraph) act faithfully in the $2^{n/2}$-dimensional
representation. We find $\upsilon_j=
(1+\gamma_j\gamma_{j+1})/\sqrt{2}$ \cite{ivanov}. It is then easy
to verify that setting $t_j=\upsilon_j$, relation
(\ref{braidrel1}) is satisfied. The center of Spin($n$) is
contained in $\widetilde{\cal B}_n^+$. It includes the elements
$U=\upsilon_1^2\upsilon_3^2\cdots
\upsilon_{n-1}^2=\gamma_1\gamma_2\cdots\gamma_n$ and
$\upsilon_j^4=-1$. For $n/2$ odd, $U^2=-1$, and $U$ generates the
center $\cong {\bf Z}_4$ of Spin($n$). The two irreducible
components are distinguished by the values $U=i$, $-i$. In these
cases, ${\bf Z}_4$ and hence the whole of $\widetilde{\cal B}_n^+$
are represented faithfully in the $2^{n/2-1}$-dimensional
representations, and hence the image of $B_n$ has order $2^n\cdot
n!$.
%(and the center of $\widetilde{\cal B}_n^+$ is $\cong {\bf
%Z}_4$ for $n/2$ even).
For $n/2$ even, $U^2=1$, and the center of
Spin($n$) is $\{1,U,-U,-1\}$. $U=1$ in one irreducible component,
$U=-1$ in the other, and the reverse for $-U$. Hence for $n\geq 8$
the image of $\widetilde{\cal B}_n^+$ (and of $B_n$) is $\cong
\widetilde{\cal B}_n^+/{\bf Z}_2$ for some ${\bf Z}_2$ in either
component, and has order $2^{n-1}\cdot n!$.
%In this case, the
%center of $\widetilde{\cal B}_n^+$ is $\cong {\bf Z}_2\times{\bf
%Z}_2$.
For $n=4$, one finds \cite{jones} that
$\upsilon_3=\upsilon_1^{-1}$, $\upsilon_1$ in the two components,
and the image of $\widetilde{\cal B}_4^+$ and $B_4$ is isomorphic
to $\widetilde{\cal B}_3^+$ ($\widetilde{\cal B}_n^+$ for $n$ odd
is defined the same way as for $n$ even) of order $2^3\cdot
3!=48$. Finally, for all even $n \geq 4$, the center of the even
part of the Clifford algebra is generated by $U$, and the center
of $\widetilde{\cal B}_n^+$ is the same as that of Spin($n$).

Our conclusion for the order of the finite group generated by the
images $\upsilon_j$ of the $t_j$'s in these irreducible
representations agrees with the analysis by Jones, who showed that
the image of $B_n$ in PU($2^{n/2-1}$) has order $2^{n-2}\cdot n!$
for $n\geq 6$, and $24$ for $n=4$ (see Thm.\ 5.2 in Ref.\
\cite{jones}). This is consistent with our results since passing
to the projective group involves division by the center (the
center of $\widetilde{\cal B}_3^+$ is ${\bf Z}_2$).

For comparison, the symmetric group $S_n$ can be viewed as the
Coxeter group ${\cal A}_{n-1}$ \cite{hump} [the Weyl group of
su($n$)]. As such it is generated by reflections (representing the
$s_j$'s) in the hyperplanes $x_j=x_{j+1}$ in ${\bf R}^n$, and this
represents it as a subgroup of O($n$). As all the generators leave
the points on the line $x_1=x_2=\cdots x_n$ fixed, the reflections
can be restricted to the orthogonal hypersurface $\sum_j x_j=0$,
and so generate a subgroup of O($n-1$). O($n-1$) has an
irreducible projective spinor representation [or linear
representation of its double cover Pin($n-1$)] of dimension
$2^{n/2-1}$, in which the lift of a reflection in any hyperplane
is represented by a linear combination of generators of a Clifford
algebra on $n-1$ generators. The lifts $\sigma_j'$, $z\sigma_j'$
to Pin($n-1$) of $s_j$ ($j=1$, \ldots, $n-1$) then generate
$\widehat{S}_n$. In terms of the Clifford algebra (for convenience
we will continue to use the Clifford algebra associated with ${\bf
R}^n$), the explicit expressions are
$\sigma_j'=(\gamma_j-\gamma_{j+1})/\sqrt{2}$ (these elements
generate a full Clifford algebra on $n-1$ generators), and the
anticommutation of $\sigma_j'$, $\sigma_k'$ for $|j-k|>1$ follows
\cite{hh}. This is {\em not} the construction proposed in Ref.\
\cite{nayak} for the braiding operations. If an abelian factor
$e^{i\theta}$ is tensored into each $\sigma_j'$, then the image of
$\widetilde{B}_n$ in U($2^{n/2-1}$) is again a finite group if
$\theta/2\pi$ is rational. Even if this finite group happens to
have the same order as $\widetilde{\cal B}_n^+$, it has a
different structure, as we have already shown.

We should mention that the statistics described by representations
of the group $\widetilde{\cal B}_n^+$ discussed here cannot
describe particles in more than two space dimensions, because the
exchanges $T_j$ do not obey (even up to a phase) the well-known
conditions $T_j^2=1$ that are required \cite{dhr,haag} in higher
dimensions.

There are also other examples of quantum Hall systems with
nonabelian braid statistics, with no obvious relation to Clifford
algebras. In the sequence of quantum Hall states, labelled by
$k=1$, $2$, \ldots, constructed in Ref.\ \cite{rr2}, the braiding
of the quasiparticles is the same as that of Wilson lines in SU(2)
Chern-Simons gauge theory of level $k$, up to tensoring by an
abelian representation. It is known that the image of the braid
group in U($N$) (for certain $N$) in these cases is finite for
$k=1$, $2$, $4$ (abelian for $k=1$), and dense in SU($N$) for all
other $k$ \cite{freed2}. Therefore in general, study of the
statistics involves the braid group, and not a finite group.

To conclude, we have pointed out that the image of the braid group
in any $2^{[(n-1)/2]}$-dimensional representation is not
isomorphic to the nontrivial double cover of the symmetric group,
even if an abelian representation of the braid group is tensored
with the latter. Projective permutation statistics is not
consistent with locality, but the physical examples in quantum
Hall states are described by the braid group and are consistent
with locality. In the case of the quasiparticles in the Moore-Read
state, the statistics is nonetheless related to Clifford algebras.

\acknowledgements

I am grateful to Zhenghan Wang and J\"urg Fr\"ohlich for helpful
communications. This work was supported by the NSF under grant
no.\ DMR-98-18259.


\vspace*{-5mm}



\begin{references}


\vspace*{-15mm}

\bibitem{mr}
G. Moore and N. Read, Nucl.\ Phys.\ B{\bf 360}, 362 (1991);
N.~Read and G.~Moore, Prog.\ Theor.\ Phys.\ (Kyoto) Supp.\ {\bf
107}, 157 (1992).

\bibitem{read00}
N. Read, Physica B {\bf 298}, 121 (2001)
[=cond-mat/0011338].

\bibitem{nayak}
C.~Nayak and F.~Wilczek, Nucl. Phys. B{\bf 479}, 529 (1996).

\bibitem{rr}
N. Read and E. Rezayi, Phys. Rev. B {\bf 54}, 16864 (1996).

\bibitem{rg}
N. Read and D. Green, \prb {\bf 61}, 10267 (2000).

\bibitem{wilczek}
F. Wilczek, hep-th/9806228.

\bibitem{bfgs}
J. Baugh, D.R. Finkelstein, A. Galiautdinov, and H. Saller, J.
Math. Phys. {\bf 42}, 1489 (2001).

\bibitem{fg}
D.R. Finkelstein and A.A. Galiautdinov, J. Math. Phys. {\bf 42},
3299 (2001).

\bibitem{g}
A.A. Galiautdinov, hep-th/0201052.

\bibitem{freed}
M.H. Freedman, A. Kitaev, M.J. Larsen, and Z. Wang,
quant-ph/0101025.

\bibitem{yswu}
Y.-S. Wu, \prl {\bf 52}, 2103 (1984).

\bibitem{weib}
C.A. Weibel, {\it Introduction to Homological Algebra} (Cambridge
University, Cambridge, 1994), Secs.\ 6.6, 6.9.

\bibitem{dijk}
R. Dijkgraaf, hep-th/9912101.

\bibitem{hh}
P.N. Hoffman and J.F. Humphreys, {\it Projective Representations
of the Symmetric Groups} (Oxford University, Oxford, 1992).

\bibitem{hamer}
M. Hamermesh, {\it Group Theory and Its Application to Physical
Problems} (Dover, New York, 1989), p. 468.


\bibitem{dhr}
S. Doplicher, R. Haag, and J.E. Roberts, Commun. Math. Phys. {\bf
13}, 1 (1969); {\bf 15}, 173 (1969); {\bf 23}, 199 (1971); {\bf
35}, 49 (1974).

\bibitem{bf}
D. Buchholz and K. Fredenhagen, Commun. Math. Phys. {\bf 84}, 1
(1982).

\bibitem{frs}
K. Fredenhagen, K.H. Rehren, and B. Schroer, Commun. Math. Phys.
{\bf 125}, 201 (1989).

\bibitem{fgm}
J. Fr\"ohlich and F. Gabbiani, Rev. Math. Phys. {\bf 2}, 251
(1990); J. Fr\"ohlich and P.A. Marchetti, Nucl. Phys. B{\bf 356},
533 (1991).

\bibitem{dr}
S. Doplicher and J.E. Roberts, Commun. Math. Phys. {\bf 131}, 51
(1990).

\bibitem{haag}
R. Haag, {\it Local Quantum Physics: Fields, Particles, Algebras}
(Springer-Verlag, New York, 1992).

\bibitem{jones}
V.F.R. Jones, in {\it Geometric Methods in Operator Algebras},
ed.\ H. Araki and E.G. Effros (Wiley, New York, 1986).

\bibitem{ivanov}
D.A. Ivanov, \prl {\bf 86}, 268 (2001).

\bibitem{hump}
J.E. Humphreys, {\it Reflection Groups and Coxeter Groups}
(Cambridge University Press, Cambridge, 1990).

\bibitem{rr2}
N. Read and E. Rezayi, Phys.\ Rev.\ {\bf B 59}, 8084 (1999).

\bibitem{freed2}
M. Freedman, M. Larsen, and Z. Wang, quant-ph/0001108;
math.GT/0103200 (see end of Sec.\ 4).

\end{references}
\end{document}